# Colloidal Lead Iodide Nanorings

Eugen Klein[1], Leonard Heymann[1],
Ana B. Hungria[2], Rostyslav Lesyuk[1,3], Christian Klinke[1,4,]*

[1] *Institute of Physical Chemistry, University of Hamburg,*
*Martin-Luther-King-Platz 6, 20146 Hamburg, Germany*
[2] *Universidad de Cádiz. Facultad de Ciencias, Campus Rio San Pedro, Cadiz 11510, Spain*
[3] *Pidstryhach Institute for applied problems of mechanics and mathematics of NAS of*
*Ukraine, Naukowa str. 3b, 79060 Lviv, Ukraine*
[4] *Department of Chemistry, Swansea University – Singleton Park,*
*Swansea SA2 8PP, United Kingdom*

**Abstract**

*Colloidal chemistry of nanomaterials experienced a tremendous development in the last decades. In the course of the journey 0D nanoparticles, 1D nanowires, and 2D nanosheets have been synthesized. They have in common to possess a simple topology. We present a colloidal synthesis strategy for lead iodide nanorings, with a non-trivial topology. First, two-dimensional structures were synthesized in nonanoic acid as the sole solvent. Subsequently, they underwent an etching process in the presence of trioctylphosphine, which determines the size of the hole in the ring structure. We propose a mechanism for the formation of lead iodide nanosheets which also explains the etching of the two-dimensional structures starting from the inside, leading to nanorings. In addition, we demonstrate a possible application of the as-prepared nanorings in photodetectors. These devices are characterized by a fast response, high gain values, and a linear relation between photocurrent and incident light power intensity over a large range. The synthesis approach allows for inexpensive large-scale production of nanorings with tunable properties.*

* Corresponding author: klinke@chemie.uni-hamburg.de

# 1. Introduction

In the realm of nanostructures new functionality comes with tuned shape and size due to quantum confinement,[1] surface plasmons or magnetic size effects.[2,3] Most of the nanomaterials produced by colloidal chemistry possess trivial topology (with genus zero) and the shapes are zero-dimensional spheres (nanocrystals),[4] one-dimensional wires,[5] or two-dimensional platelets.[6] A rare exception with a new topology are nanocages.[7] Such nanomaterials, in particular 2D semiconductor nanosheets, may be employed in high performance field-effect transistors (FETs),[8,9,10] photodetectors,[11,12] photovoltaic devices,[13,14] and diodes.[15]

In order to construct nanosheet-based devices, shape, size and composition control is essential. Various techniques like exfoliation of layered structures,[16,17] solvothermal methods,[18,19] chemical vapor deposition,[20,21] and direct colloidal synthesis in a flask have proven to be effective.[22,23] However, it is still challenging to prepare uniform and single crystalline sheets without any domain formation in the whole sheet. A way to increase the chances to obtain such well-defined structures is to use layered systems which have a strong intralayer but a weak layer to layer chemical bonding.[24] These materials tend to crystallize in two-dimensional structures at lenient conditions and do not need to be forced to by high temperatures or strong ligands.[25,26]

Ring-like structures, which represent a new topology, become increasingly popular. They attracted already significant attention due to their unique optical, electronic, magnetic and catalytic properties.[27-31] For example, Pd nanorings showed superior catalytic activity toward hydrogenation compared to Pd nanosheets as well as commercial Pd black. They are also highly stable during this process.[32] Cobalt or nickel magnetic nanoring arrays may find potential application in magnetic storage devices.[30,33,34] Other important examples are gold or silver nanorings with applications in biosensors and InGaAs or GaN semiconductor nanorings in optoelectronical devices.[35-38] Potentially, the colloidal nano and microrings could be used for lab-on-a-chip systems,[39] ring resonators[40] and sensors, as unique diffraction templates and building blocks for hybrid devices based on 2D materials. An interesting feature is the tunability of the optical properties of these structures. Therefore, having control over the lateral dimensions and thicknesses of these rings is important.

The most common way to prepare ring structures is colloidal lithography.[28,41] This technique and its variations, e.g. lithographically patterned nanoscale electrodeposition (LPNE),[42,43] are tools for patterning planar surfaces with arrays of metal and soft matter rings. Other successful techniques for preparing nanorings of various materials are solvothermal approaches and oxidative etching of as prepared nanosheets with halogen ions.[32,44,45] The large-scale synthesis of nanorings with lateral extensions exceeding 100 nm is rather difficult, though it would allow the measurement of the characteristics of individual crystals, and their efficient application in optoelectronic devices.

Lead iodide is a direct band gap semiconductor with a gap between 2.3 and 2.4 eV.[46] The crystal structure consists of layers of hexagonally close packed iodine and lead atoms, oriented perpendicular to the c-axis.[24,47,48] The potential applications for this material are high energy photon detectors for X-rays and gamma rays and photocells.[24] Lead iodide finds also application as precursor materials in the perovskite solar cell fabrication.[49]

We report the synthesis and characterization of $PbI_2$ nanorings prepared via a direct colloidal route. The ring structures are obtained by etching as prepared $PbI_2$ nanosheets with trioctylphosphine (TOP). These nanosheets were prepared following the synthetic route described in our recent paper by changing the solvent from oleic acid to nonanoic acid.[50] The nanorings are analyzed by TEM, XRD, AFM and UV/Vis spectroscopy techniques. To our best knowledge, we report for the first time syntheses of $PbI_2$ rings. They possess thicknesses between 20 nm and 85 nm and lateral dimensions of up to 10 μm. Additionally to the possible formation process, we also provide electrical and optical measurements of single crystal nanorings and their application in photodetector devices.

## 2. Results and Discussion

### 2.1. Characterization of the ring structures

Figure 1 shows ring structures of $PbI_2$, which were prepared by etching of corresponding nanosheets using TOP. The $PbI_2$ nanosheets were synthesized in nonanoic acid.[50] At the end of the reaction these sheets were centrifuged once to separate them from the reaction mixture. Subsequently, they were re-suspended in toluene and stored in a freezer. These nanosheets have a uniform hexagonal shape, sizes of 2 μm to 10 μm and a thickness of 20 nm (Figure 1A). For the synthesis of rings, a part of these sheets was mixed with diphenylether (DPE) and a small amount of trioctylphosphine (TOP). Then, the mixture was heated to the desired temperature of 30 °C to 250 °C for 10 min. Figure 1B shows thin ring structures, which were synthesized at 100 °C with lateral sizes of 2 µm to 10 µm. These structures have a strong tendency to stack with the same orientation. Further, they possess a centered hole and frayed edges. Choosing a reaction temperature of 130 °C yields thicker rings with a centered hole of the same size (Figure 1C). Figures 1D and E depict rings synthesized at 170 °C and 200 °C, respectively. These structures have a smoother surface and well-defined edges compared to the rings prepared at lower temperatures. Additionally, the structures at 200 °C show the largest thickness. The temperature of the etching process determines the thickness of the ring structures. Figure 1F depicts selected area electron diffraction (SAED) of all ring structures, which show dot patterns in all cases. The thickness of these rings was measured by AFM and calculated by the Scherrer equation from the data of the corresponding XRD patterns shown in Figure 2. The thickest rings possess a thickness of 85 nm measured by AFM and 75 nm measured by XRD whereas the thinnest rings show a thickness of only 25 nm by AFM and 21 nm by XRD. TEM tomography confirms the morphology of the structures to be rings with lateral dimensions in the micrometer range and heights of a few tens of nanometers (see the videos in the Supporting Information). Additionally, energy-dispersive X-ray spectroscopy (XEDS) elemental mapping shows the spatial distribution of Pb, I, and Pb+I of the set of nanorings. A high-angle annular dark-field scanning transmission electron microscope (HAADF-STEM) image is presented for comparison (Figure S1).

The crystal structure of the thinnest and the thickest rings is hexagonal P-3m1 2H type. Syntheses over 200 °C lead to mainly destroyed ring structures (Figure S2A). Rings prepared at temperatures below 100 °C possess thicknesses between 20 nm and 30 nm. Their morphology is similar to the one of products prepared at 100 °C, although a considerable amount of by-products is present in the shape of pieces of frayed nanosheets (Figure S2B and C). Disregarding the by-products, it is even possible to perform the etching process by simply mixing the nanosheets with TOP in a falcon tube at room temperature (Figure S2D).

The UV/Vis spectra of rings with thicknesses of 21 nm and 82 nm are similar and show a shoulder at 500 nm for absorption and near 521 nm for emission (Figure 3 A, B). With the help of the Tauc linearization method for the absorption edge the optical band gap of 2.41 and 2.40 eV for the thin and thick nanorings was estimated (inset of the Figure 3B). We note that the average size of the nanocrystals in analyzed samples (20 and 80 nm) excludes the confinement regime, thus the slight difference in the optical band gap should not be attributed to the size effect. In Figure 3 C and D corresponding PL mapping images are presented for the individual nanorings of 21 nm and 82 nm thickness obtained by confocal excitation microscopy showing the spatial distribution of PL, which clearly corresponds to the ring-like structure. Analysis of the PL spectrum obtained from the solution of above mentioned samples shows that the emission profiles have a complex shape (Figure S3 A, B for better spectral resolution and numerical fitting). It consists of a main slightly asymmetric central peak accompanied by red-shifted side peaks. We note that despite the position of the PL maximum is identical, the spectral broadening (FWHM) differs for thin and thick rings samples being equal to nearly 81 and 52 meV (~3 and ~2 kT). In general, the PL of lead iodide may originate from free and bound-exciton transitions as well as from donor-acceptor pairs (DAP) and surfaces traps-related transitions[51,52,53]. Based on literature data, we assume that at room temperature the radiative excitonic recombination produces very weak intensity, which cannot be observed in our measurements. Taking into account the different broadening of the PL for thin and thick rings, we conclude that this difference must be attributed to the volumetric properties of the nanocrystals (such as crystal defects). The tiny Stokes-shift of nearly 30 meV does not support the assumption that the observed emission stems from the DAP-related recombination, since in this case the energy position of possible donors and acceptors in $PbI_2$ are deeper as required for the explanation.[54] We assume that observed PL might originate from the band-edge recombination of electron-hole pairs. The redundant and different broadening of the PL spectrum for thin and thick nanorings may be explained by a slight self-doping effect due to the presence of defects. Since the thick nanorings are prepared by sufficiently larger temperature compared to the thin ones, they are more homogeneous and have better crystallinity, thus the self-doping effect might be reduced. Therefore the optical band gap slightly decreases for thicker nanorings. The lower-energy shoulders can originate from mentioned DAP-related recombination and surface-related traps[51], however to make unambiguous assignments further studies are needed. It can be additionally seen (Figure 3 C, D) that the highest emission intensities can be observed at the corner of the hexagonal nanorings, whereas no emission can be observed in the center. The emission in the areas that lie in between is a little lower compared to the edges. We explain this by the lower crystal quality in the center of the structure, giving rise to the defect-related non-radiative relaxation processes for excited carriers. This is further explained below in the discussion of the formation process. Figures S3 C and D show a TEM image and emission measured by confocal microscopy for non-etched $PbI_2$ nanosheets. The emission shows a homogeneous distribution over the whole lateral size with the exception of the middle and two spots between the edges and the middle. TEM images of the same samples show no signs of holes for the nanosheets. Differences in the PL spectrum across the rings or nanosheets were not detected, probably due to the low quantum yield of the sample and instrumental restrictions.

In series of test experiments we studied the role of TOP in the etching process. The sample of PbI$_2$ nanosheets was thoroughly washed and divided into three parts. One was diluted in toluene with a small amount of oleic acid, one with TOP and one was left in toluene as prepared for comparison. The untreated sheets and the ones suspended with oleic acid and stored for three days in the freezer remained monocrystalline and undamaged. In contrast to this, the structures which were diluted with TOP underwent a considerable change in shape. All sheets had a hole in the middle, were deformed in their usual hexagonal shape and showed rough edges (corresponding TEM images are presented in Figure S4). Further we performed standard PbI$_2$ nanosheet synthesis reaction for a longer time or with a higher amount of TOP as usually to test whether ring structures could be obtained in a one pot approach. However, the only product present even after one hour was monocrystalline nanosheets.

Being responsible for the hole formation, TOP in different amounts can also influence the size of the hole of the ring structures. Figure S5 shows TEM images of four samples prepared at 100 °C and 200 °C with different concentrations of TOP. For 100 °C the size of the hole changes from about 100 nm prepared with 0.08 mL of TOP to 1.5 µm prepared with 0.12 mL. At 200 °C the size changes from about 400 nm to 1.5 µm, synthesized with 0.06 mL and 0.08 mL of TOP, respectively.

## 2.2. Formation of ring structures

In order to understand the ring formation for PbI$_2$, aliquots were taken during the synthesis of PbI$_2$ nanosheets performed in oleic acid and nonanoic acid. Figure 4 depicts TEM images at two stages in the formation of nanosheets in oleic acid. The first image indicates an agglomeration of particles to a sheet-like shape while the structure is partly crystallized. The second image shows monocrystalline sheets (SAED of both samples presented in the Supporting Information S6). Scheme 1 presents a mechanism for these observations where at first small particles are formed, followed by agglomeration and further crystallization with time to form a single crystal. Figure 5 shows the formation of nanosheets in nonanoic acid. First, a cloud of nonanoic acid and lead nonanoate forms (Figure 5A) which develops to a ring shape (Figure 5B). Figures 5C and D show an agglomeration of particles, which crystallize at the edges of ring or half ring structures. Figures 5E and F show the further growth of the structures, which continues as shown in Figure 5G, where a sheet with a small hole in the middle is displayed. The aliquots were taken every five seconds. At the end of the reaction, every hole is closed and the sample consists of monocrystalline PbI$_2$ nanosheets depicted in Figure 5H (SAED and an overview of nanosheets at the end of the reaction in S6). The 2D arrangement of the micelles is conditioned by the formation of lamellar soft templates where the hydrophilic lead ion layers are separated by bilayers of the corresponding carboxylic acids.[55,56]

Scheme 2A presents a mechanism for these observations where at first a micelle of lead nonanoate molecules together with nonanoic acid molecules is formed. In order to increase the concentration of lead nonanoate in the micelle a separation of the two present species occurs. This step is crucial for the formation of two-dimensional lead iodide nuclei. The as-formed micelle changes into a ring or half-ring shape due to differences in the polarity of the lead nonanoate molecules and nonanoic acid, the steric hindrance of these molecules and concentration alterations. The separation in form of pushing all of the acid to the outside of the micelle cannot be completely achieved because of the large size of the micelle. Therefore, the acid is pushed in two directions, towards the outside and towards the center of the micelle. The inside of the ring consists of ordered lead nonanoate molecules, which build a lamellar structure. The outside and inside interfaces are composed of nonanoic acid solely (Figure 6A and C). Figure 6B shows a large sheet after the etching process where the decomposition started not only in the center but also at many other regions of the structure. It seems that in this case, the micelle was too large and the nonanoic acid was pushed at several areas in the structure. If oleic acid is used instead of nonanoic acid the separation seems to be insufficient and the micelle remains as it was. Oleic acid has a two times longer alkyl chain, which leads to a very slow diffusion through the micelle compared to nonanoic acid. Therefore, oleic acid is not able to leave the micelle and slows down the formation of lead iodide monomers and their agglomeration to nuclei, which results in a two times longer reaction time for the synthesis in oleic acid compared to nonanoic acid. After the micelle changed its shape PbI$_2$ particles were formed in the ring-like structure. In Scheme 2A step 3 the particles crystallize at the edges of the ring-like structure first, following a further growth to complete the ring. At step 5 only a small hole is present which vanishes with time completing a PbI$_2$ nanosheet. Scheme 2B presents the etching of the PbI$_2$ nanosheets prepared in nonanoic acid. Thus, the center of these sheets is formed at the end of the synthesis. Many crystallographic defects can be present, like planar defects or bulk defects. TOP functioned as the etching reactant, which starts to decompose the sheet from the middle since this is the last area where the sheet crystallizes and therefore the richest one in defects. Depending on the temperature or how much TOP is used the decomposition occurs faster or slower and it is possible to prepare rings with different thicknesses. Performing the etching process on nanosheets, which were prepared in octanoic acid, heptanoic acid or hexanoic acid yields similar ring structures like for nonanoic acid (Figure S7). In contrast to this when we perform the etching process on sheets prepared in oleic acid the decomposition starts everywhere at the same time and the structure falls apart (Figure S8).

## 2.3. Electrical characterization

Furthermore, the photoelectrical properties of the as-prepared individual nanorings of two different thicknesses of 21 nm and 82 nm have been measured and characterized. In Figure 7 A the IV curves for a device made of the thicker ones are presented on a logarithmic and a linear scale under illumination with a 405 nm laser diode. The IV curves show an S-shape with a small plateau around zero bias, which is typical for Schottky contacts. An AFM image of the device is shown in Figure 7D. The responsivity *R*, the external quantum efficiency (*EQE)* as well as the response time and the specific detectivity (*D\*)* were determined and are presented in Table 1. It was found that for the thicker nanorings better results in term of the observed currents as well as the above-mentioned figure of merits (FOM) for photodetectors could be achieved. We ascribed that to the crystal quality, which appears to be influenced by an increased number of defects at smaller thicknesses due to the above-described etching process. Nevertheless, the nanoring devices exhibit improved properties compared to devices based on nanosheets. The nanosheet devices show a very low response to illumination and the resulting currents were too low for further characterization as photodetector devices. The better properties of the nanorings devices can be explained by the improved crystal quality of the nanorings due to longer reaction and annealing times. All the nanoring devices measured were defined by a dark current at the noise level of $8 \cdot 10^{-15}$ A. Therefore, large on/off-ratios (gain) of up to $1.14 \cdot 10^5$ can be observed at 10 V. This parameter is important for optoelectronic applications like the conversion of an optical into an electrical signal. For these applications it is also important to have a fast response time, which can be differentiated into rise and fall time. The rise time is defined as the time needed for the current to rise to 90 % of the maximum value, whereas the decay to 10 % of the maximum values defines the fall time. For the fastest device we observed 2.5 and 3.9 ms for rise and fall time, respectively (Figure 7B). The responsivity was calculated by $R=I/P \cdot A$ and the external quantum efficiency by $EQE=hc \cdot R/e \cdot \lambda$, where *I* is the photoexcited current, *P* is the light power intensity, *A* the area of the photodetector, $h$ is Planck's constant, $c$ is speed of light, $e$ is electron charge and $\lambda$ is the excitation wavelength. Despite the comparable small responsivity and external quantum efficiency the specific detectivity ($D^*=R \cdot A^{1/2}/(2e \cdot I_{dark})$) is in a reasonable dimension due to the large gain values resulting from the dark current $I_{dark}$ in the noise level. Figure 7C shows the linear dependency of the light power intensity and the resulting photocurrent over a large power range (1.53 mW/cm² to 315 mW/cm²). This property is important for photosensor applications, which would be limited by saturation at higher intensities. The corresponding images for the devices made of thinner nanorings are shown in Figure S9. The evaluated FOM are presented in Table 1 as well as the selected data from the literature for comparison.[57,58,59] The PbI$_2$ nanorings are characterized by a fast response and high gain values, but lack in responsivity and external quantum efficiency compared to other materials prepared by colloidal synthesis. The specific detectivity is in a reasonable range. If compared to graphene the responsivity as well as the external quantum efficiency is in the same range, whereas the gain value is at least 3 orders of magnitude larger. The here described PbI$_2$ nanorings are suited for inexpensive high-energy detectors because of the large absorption cross-section resulting from the heavy element lead.

## 3. Conclusion

In summary, we described the synthesis of 2D nanorings of PbI$_2$ mediated by an etching process of as prepared nanosheets. The nanosheets were synthesized using a colloidal route. The nanorings were prepared by an etching process and characterized by TEM, XRD, SAED, AFM, STEM XEDS and tomography techniques and optical spectroscopy. By varying the temperature of the etching process, rings with thicknesses between 20 nm and 85 nm could be achieved. Variation of the TOP amount leads to rings with tunable inner diameters between 100 nm and 1.5 µm. The outer diameters of all ring structures were between 2 µm and 10 µm, predetermined by the nanosheet synthesis conditions. Additionally, we investigated the formation of these nanosheets and nanorings by taking aliquots during the corresponding reactions. We found that for the nanosheets prepared in nonanoic acid micelles were present at an early stage of the reaction, which transform to rings. The formation of the clouds occurs due to polarity differences of the carboxylic acids and their corresponding deprotonated lead molecules and steric differences for the different acids. These clouds are the templates for the nanosheets, which grow as rings and half rings to ultimately form complete nanosheets. Therefore, the etching with TOP starts from the center of these nanosheets since this is the point which crystallized last and it is the weakest area in the structure. In case of oleic acid the formation takes place everywhere at the same time and the etching decomposes the structures from the outside as well as from the inside. Further, the application of the as-prepared nanorings in photodetection devices has been demonstrated. The rings exhibit a fast response time in the lower milliseconds range and are characterized by high gain values ($10^5$) as well as a linear light power intensity to photocurrent relation.

## 4. Experimental Section

*Materials and Equipment:* All chemicals were used as received: Lead(II) acetate tri-hydrate (Aldrich, 99.999%), oleic acid (OA, Aldrich, 90%), nonanoic acid (Alfa Aesar, 97%), octanoic acid (Aldrich, >99%), heptanoic acid (Aldrich, >99%), hexanoic acid (Aldrich, >99%), tri-octylphosphine (TOP; ABCR, 97%), 1,2-diiodoethane (DIE; Aldrich, 99%), poly (methyl methacrylate) in anisole (Allresist, AR-P 632.06), methyl isobutyl ketone (MIBK, Aldrich, >99%), 2-propanol (ACS, >99.7), and chlorobenzene (Aldrich, >99.5%).

*$PbI_2$ nanosheet synthesis:* In a typical synthesis a three neck 50 mL flask was used with a condenser, septum and thermocouple. 860 mg of lead acetate tri-hydrate (2.3 mmol) were dissolved in 20 mL of nonanoic acid (114 mmol) and heated to 80 °C until the solution turned clear in a nitrogen atmosphere. Then vacuum was applied to remove the acetic acid which is generated by the reaction of nonanoic acid with the acetate from the lead precursor. After 1.5 h the reaction apparatus was filled with nitrogen again and 2 mL of a 48.7 mg 1,2-diiodoethane (0.17 mmol) in 3 mL nonanoic acid precursor was added at 80 °C to the solution. After 4 min 0.06 mL of tri-octylphosphine (0.13 mmol) was added to the reaction solution. After 2.5 min the heat source was removed and the solution was left to cool down to below 60 °C. Afterwards, it was centrifuged at 4000 rpm (3756 rcf) for 3 min. The particles were suspended in 12 mL toluene and placed into a freezer for storage.

*Synthesis with other carboxylic acids:* The procedure was similar to the nonanoic acid synthesis with just different carboxylic acids used as the main ligand and for the preparation of the DIE precursors: with oleic acid: 20 mL (60 mmol), with octanoic acid: 20 mL (126 mmol), with heptanoic acid: 20 mL (140 mmol), and with hexanoic acid: 20 mL (160 mmol).

*$PbI_2$ nanoring synthesis:* A three neck 50 mL flask was used with a condenser, septum and thermocouple. 10 mL of diphenyl ether (63.5 mmol) and an amount of TOP between 0.06 mL and 1.2 mL (0.13 mmol – 0.27 mmol) were heated to 80 °C in a nitrogen atmosphere. Then vacuum was applied to dry the solution. After 1 h the reaction apparatus was filled with nitrogen again and heated to the desired temperature between 30 °C and 250 °C. The synthesis was started with the injection of 1 mL of the as prepared $PbI_2$ nanosheet in toluene. After 10 min the heat source was removed and the solution was left to cool down below 60 °C. Afterwards, it was centrifuged at 4000 rpm (3756 rcf) for 3 min. The particles were washed two times in toluene before the product was finally suspended in toluene again and put into a freezer for storage.

The TEM samples were prepared by diluting the nanostructure suspensions with toluene followed by drop casting 10 µL of the suspension on a TEM copper grid coated with a carbon film. Standard images were done on a JEOL-1011 with a thermal emitter operated at an acceleration voltage of 100 kV. The compositional maps were performed by X-ray Energy Dispersive Spectroscopy in STEM mode (STEM-XEDS) using a double aberration-corrected FEI Titan3 Themis 60–300 microscope equipped with a high efficiency, high sensitivity, 4-detector ChemiStem system. Very high spatial resolution STEM-XEDS maps were acquired using a high brightness, sub-angstrom (0.07 nm) diameter, electron probe in combination with a highly stable stage which minimized sample drift. Element maps were acquired with a screen current of 80-120 pA and a pixel time of 13 ms which results in a total acquisition time of approximately 40 minutes. An averaging filter was used on the images as provided in the Esprit software from Bruker. The FEI Titan3 Themis 60-300 double aberration corrected microscope was used at 200 kV to acquire the STEM tomography tilt series. A convergence angle of 9 mrad was selected in order to improve the depth of focus and a camera length of 115 mm was used. The software FEI Explore3D v.4.1.2 facilitated the acquisition of the tomography tilt series from −56° to +62° every 2° and the alignment and reconstruction of the data set. Avizo software was used for visualization. X-ray diffraction (XRD) measurements were performed on a Philips X'Pert System with a Bragg-Brentano geometry and a copper anode with a X-ray wavelength of 0.154 nm. The samples were measured by drop-casting the suspended nanostructures on a <911> or <711> cut silicon substrate. Atomic force microscopy (AFM) measurements were performed in tapping mode on a JPK Nano Wizard 3 AFM in contact mode. Images were taken of the as-prepared nanoring devices. UV/vis absorption spectra were obtained with a Cary 5000 spectrophotometer equipped with an integration sphere. The PL spectra measurements were obtained by a fluorescence spectrometer (Fluoromax-4, Horiba Jobin Yvon).

*Electrical Characterization:* 100 µL of the solution of the $PbI_2$ nanorings was diluted with 500 µL of toluene and dropcasted on a doped silicon wafer of roughly 1 cm² and 300 nm of thermal oxide. This wafer was covered with marker fields, which were prepared in the same way as the contacts. The procedure is described in the following. The wafer was covered with the positive resist poly (methyl methacrylate) (PMMA) in anisole via spin-coating (60 s, 4000 rpm). Standard electron-beam lithography with a Quanta Scanning Electron Microscope (FEI) and CAD software ELPHY Plus (Raith) was followed by developing (60 s, 12.5 % methyl isobutyl ketone in 2-propanol) the structures and metal evaporation of 3 nm titanium as an adhesion layer and 22 nm of gold as electrode material. As a lift-off reagent chlorobenzene was used. The electrical characterizations were carried out in a Lakeshore-Desert vacuum probe station (residual gas pressure of $10^{-5}$ mbar) equipped with a Keithley 4200-SCS parameter analyzer. For illumination, Cobolt 06-MLD laser diode was used and modulated by an external function generator (LEADER LFG-1300). The total number of measured devices was 15 for both thicknesses.

**Supporting Information**

Supporting Information is available from the Wiley Online Library or from the author.

**Acknowledgments**

The authors thank the Alf Mews group for providing the Confocal Microscopy setup. Further, the German Research Foundation DFG is acknowledged for financial support in the frame of the Cluster of Excellence "Center of ultrafast imaging CUI" and for granting the project KL 1453/9-2. The European Research Council is acknowledged for funding an ERC Starting Grant (Project: 2D-SYNETRA (304980), Seventh Framework Program FP7). We further acknowledge MINECO (Spain) for the project MAT2016-81118-P.

**Conflict of Interest**

The authors declare no conflict of interest.

**References**

1 C. B. Murray, D. J. Noms, M. G. Bawendi, *J. Am. Chem. Soc.* 1993, **115**, 8706-8715.

2 M. Hojeij, N. Younan, L. Ribeaucourt, H. H. Girault, *Nanoscale* 2010, **2**, 1665-1669.

3 J. Rivas, M. Banobre-Lopez, Y. Pineiro-Redondo, B. Rivas, M. A. Lopez-Quintela, *J. Magn. Magn. Mater.* 2012, **324**, 3499-3502.

4 Z. Quan, Z. Luo, Y. Wang, H. Xu, C. Wang, Z. Wang, J. Fang, *Nano Lett.* 2013, **13**, 3729-3735.

5 K. S. Cho, D. V. Talapin, W. Gaschler, C. B. Murray, *J. Am. Chem. Soc.* 2005, **127**, 7140-7147.

6 T. Bielewicz, M. M. R. Moayed, V. Lebedeva, C. Strelow, A. Rieckmann, C. Klinke, *Chem. Mater.* 2015, **27**, 8248-8254.

7 S. E. Skrabalak, J. Chen, Y. Sun, X. Lu, L. Au, C. M. Cobley, Y. Xia, *Acc. Chem. Res.* 2008, **41**, 1587-1595.

8 S. Dogan, T. Bielewicz, Y. Cai, C. Klinke, *Appl. Phys. Lett.* 2012, **101**, 073102.

9 Y. T. Lee, W. K. Choi, D. K. Hwang, *Appl. Phys. Lett.* 2016, **108**, 253105.

10 P. Agnihotri, P. Dhakras, L. U. Lee, *Nano Lett.* 2016, **16**, 4355-4360.

11 R. Velazquez, A. Aldalbahi, M. Rivera, P. Feng, *AIP Advances* 2016, **6**, 085117.

12 S. Oh, J. Kim, F. Ren, S. J. Pearton, J. Kim, *J. Mater. Chem.* 2016, **10**, 1039.

13 S. Dogan, T. Bielewicz, V. Lebedeva, C. Klinke, *Nanoscale* 2015, **7**, 4875.

14 T. Bielewicz, S. Dogan, C. Klinke, *Small* 2015, **11**, 826.

15 E. Lhuillier, S. Pedetti, S. Ithurria, B. Nadal, H. Heuclin, B. Dubertret, *Acc. Chem. Res.* 2015, **48**, 22-30.

16 J. N. Coleman, M. Lotya, A. O'Neill, S. D. Bergin, P. J. King, U. Khan, K. Young, A. Gaucher, S. De, R. J. Smith, I. V. Shvets, S. K. Arora, G. Stanton, H. Y. Kim, K. Lee, G. T. Kim, G. S. Duesberg, T. Hallam, J. J. Boland, J. J. Wang, J. F. Donegan, J. C. Grunlan, G. Moriarty, A. Shmeliov, R. J. Nicholls, J. M. Perkins, E. M. Grieveson, K. Theuwissen, D. W.

McComb, P. D. Nellist, V. Nicolosi, *Science* 2011, **331**, 568-571.

17 R. Ma, T. Sasaki, *Adv. Mater.* 2010, **22**, 5082-5104.

18 Z. Cui, L. Mi, D. Zeng, *J. Alloys Compd.* 2013, **549**, 70-76.

19 M. Safdar, Z. Wang, M. Mirza, F. K. Butt, Y. Wang, L. Sun, J. He, *J. Mater. Chem. A* 2013, **1**, 1427-1432.

20 J. C. Shaw, H. Zhou, Y. Chen, N. O. Weiss, Y. Liu, Y. Huang, X. Duan, *Nano Res.* 2014, **7**, 511.

21 K. Xu, Z. Wang, X. Du, M. Safdar, C. Jiang, J. He, *Nanotechnology* 2013, **24**, 465705.

22 S. Lee, D. T. Lee, J. H. Ko, W. J. Kim, J. Joo, S. Jeong, J. A. McGuire, Y. H. Kim, D. C. Lee, *RSC Adv.* 2014, **4**, 9842-9850.

23 T. Bielewicz, E. Klein, C. Klinke, *Nanotechnology* 2016, **27**, 355602.

24 Y. C. Chang, R. B. James, *Phys. Rev. B* 1997, **55**, 8219.

25 F. Chen, L. Wang, X. Ji, Q. Zhang, *ACS Appl. Mater. Interfaces* 2017, **9**, 30821-30831.

26 C. Tan, H. Zhang, *Nat. Commun.* 2015, **6**, 1-13.

27 P. J. Pauzauskie, D. J. Sirbuly, P. Yang, *Phys. Rev. Lett.* 2006, **69**, 143903-1 – 143903-4.

28 J. Aizpurua, P. Hanarp, D. S. Sutherland, M. Käll, G. W. Bryant, F. J. Garcia de Abajo, *Phys. Rev. Lett.* 2003, **90**, 057401-1 - 057401-4.

29 F. Pederiva, A. Emperador, E. Lipparini, *Phys. Rev. B* 2002, **66**, 165314-1 - 165314-6.

30 Z. K. Wang, H. S. Lim, H. Y. Liu, S. C. Ng, M. H. Kuok, L. L. Tay, D. J. Lockwood, M. G. Cottam, K. L. Hobbs, P. R. Larson, *Phys. Rev. Lett.* 2005, **94**, 137208.

31 C. Chen, Y. Kang, Z. Huo, Z. Zhu, W. Huang, H. L. Xin, J. D. Snyder, D. Li, J. A. Herron, M. Mavrikakis, M. Chi, K. L. More, Y. Li, N. M. Markovic, G. A. Somorjai, P. Yang, V. R. Stamenkovic, *Science* 2014, **343**, 1339-1343.

32 Y. Li, W. Wang, K. Xia, W. Zhang, Y. Jiang, Y. Zeng, H. Zhang, C. Jin, Z. Zhang, D. Yang, *Small* 2015, **11**, 4745-4752.

33 Y. L. Li, S. L. Tang, W. B. Xia, L. Y. Chen, Y. Wang, T. Tang, Y. W. Du, *Appl. Phys.*

*Lett.* 2012, **100**, 183101-1 - 183101-4.

34 G. Duan, W. Cai, Y. Luo, Z. Li, Y. Lei, *J. Phys. Chem. B* 2006, **110**, 15729-15733.

35 E. M. Larsson, J. Alegret, M. Käll, D. S. Sutherland, *Nano Lett.* 2007, **7**, 1256-1263.

36 A. W. Clark, A. Glidle, D. R. S. Cumming, J. M. Cooper, *J. Am. Chem. Soc.* 2009, **131**, 17615-17619.

37 A. Lorke, R. J. Luyken, A. O. Govorov, J. P. Kotthaus, J. M. Garcia, P. M. Petroff, *Phys. Rev. Lett.* 2000, **84**, 2223-2226.

38 K. H. Li, Z. Ma, H. W. Choi, *Appl. Phys. Lett.* 2011, **98**, 071106-1 - 071106-3.

39 M. Medina-Sanchez, S. Miserere, A. Merkoci, *Lab Chip* 2012, **12**, 1932-1943.

40 N. Gaponik, Y. P. Rakovich, M. Gerlach, J. F. Donegan, D. Savateeva, A. L. Rogach, *Nanoscale Res. Lett*. 2006, **1**, 68-73.

41 M. Winzer, M. Kleiber, N. Dix, R. Wiesendanger, *Appl. Phys. A* 1996, **63**, 617 - 619.

42 E. J. Menke, M. A. Thomson, C. Xiang, L. C. Yang, R. M. Penner, *Nat. Mater.* 2006, **5**, 914-919.

43 C. Xiang, S. C. Kung, K. D. Taggart, F. Yang, M. A. Thompson, A. G. Güell, Y. Yang, R. M. Penner, *ACS Nano* 2008*,* **2***, 1939-1949*.

44 C. Peng, L. Gao, S. Yang, *Chem. Commun.* 2007, **0**, 4372-4374.

45 X. Hu, J. C. Yu, J. Gong, Q. Li, G. Li, *Adv. Mater.* 2007*,* **19***, 2324-2329*.

46 A. Ferreira da Silva, N. Veissid, C. Y. An, I. Pepe, N. Barros de Oliveira, A. V. Batista da Silva, *Appl. Phys. Lett.* 1996, **69**, 1930-1932.

47 M. Y. Khilji, W. F. Sherman, G. R. Wilkinson, *J. Raman Spectrosc.* 1982, **13**, 127.

48 R. S. Mitchell, *Z. Kristallogr.-Cryst. Mater.* 1959, **111.1-6**, 372-384.

49 L. Protesescu, S. Yakunin, M. I. Bodnarchuk, F. Krieg, R. Caputo, C. H. Hendon, R. X. Yang, A. Walsh, M. V. Kovalenko, *NanoLett.* 2015*,* **15***,* 3692-3696.

50 E. Klein, R. Lesyuk, C. Klinke, *Nanoscale* 2018, **10**, 4442-4451.

51 M. Baibarac, N. Preda, L. Mihut, I. Baltog, S. Lefrant, J. Y. Mevellec, *J. Phys. Condens. Matter* 2004, **16**, 2345-2356.

52 J. F. Condeles, R. A. Ando, M. Mulato, *J. Mater. Sci* 2008, **43.2**, 525-529.

53 S. S. Novosad, I. S. Novosad, I. M. Matviishin, *Inorganic materials* 2002, **38**, 1058-1062.

54 V. A. Bibik, N. A. Davydova, *Phys. Status Solidi (a)* 1991, **126(2)**, K191-K196.

55 P. J. Morrison, R. A. Loomis, W. E. Buhro, *Chem. Mater.* 2014, *26*, 5012-5019.

56 W. Bryks, M. Wette, N. Velez, S. W. Hsu, A. R. Tao, *J. Am. Chem. Soc.* 2014, *136*, 6175-6178.

57 S. Acharya, M. Dutta, S. Sarkar, D. Basak, S. Chakraborty, N. Pradhan, *Chem. Mater.* 2012, **24**, 1779-1785.

58 L. Lv, Y. Xu, H. Fang, W. Luo, F. Xu, L. Liu, B. Wang, X. Zhang, D. Yang, W. Hu, A. Dong, *Nanoscale* 2016, **8**, 13589-13596.

59 P. Ramasamy, D. Kwak, D. H. Lim, H. S. Ra, J. S. Lee, *J. Mater. Chem. C* 2016, **4**, 479-485.

60 F. Xia, T. Mueller, Y. M. Lin, A. Valdes-Garcia, P. Avouris, *Nat. Nanotechnol.* 2009, **4**, 839−843.

61 R. Frisenda, J. O. Island, J. L. Lado, E. Giovanelli, P. Gant, P. Nagler, S. Bange, J. M. Lupton, C. Schüller, A. J. Molina-Mendoza, L. Aballe, M. Foerster, T. Korn, M. A. Niño, D. Perez de Lara, E. M. Pérez, J. Fernandéz-Rossier, A. Castellanos-Gomez, *Nanotechnol.* 2017, **28**, 455703.

**Figures**

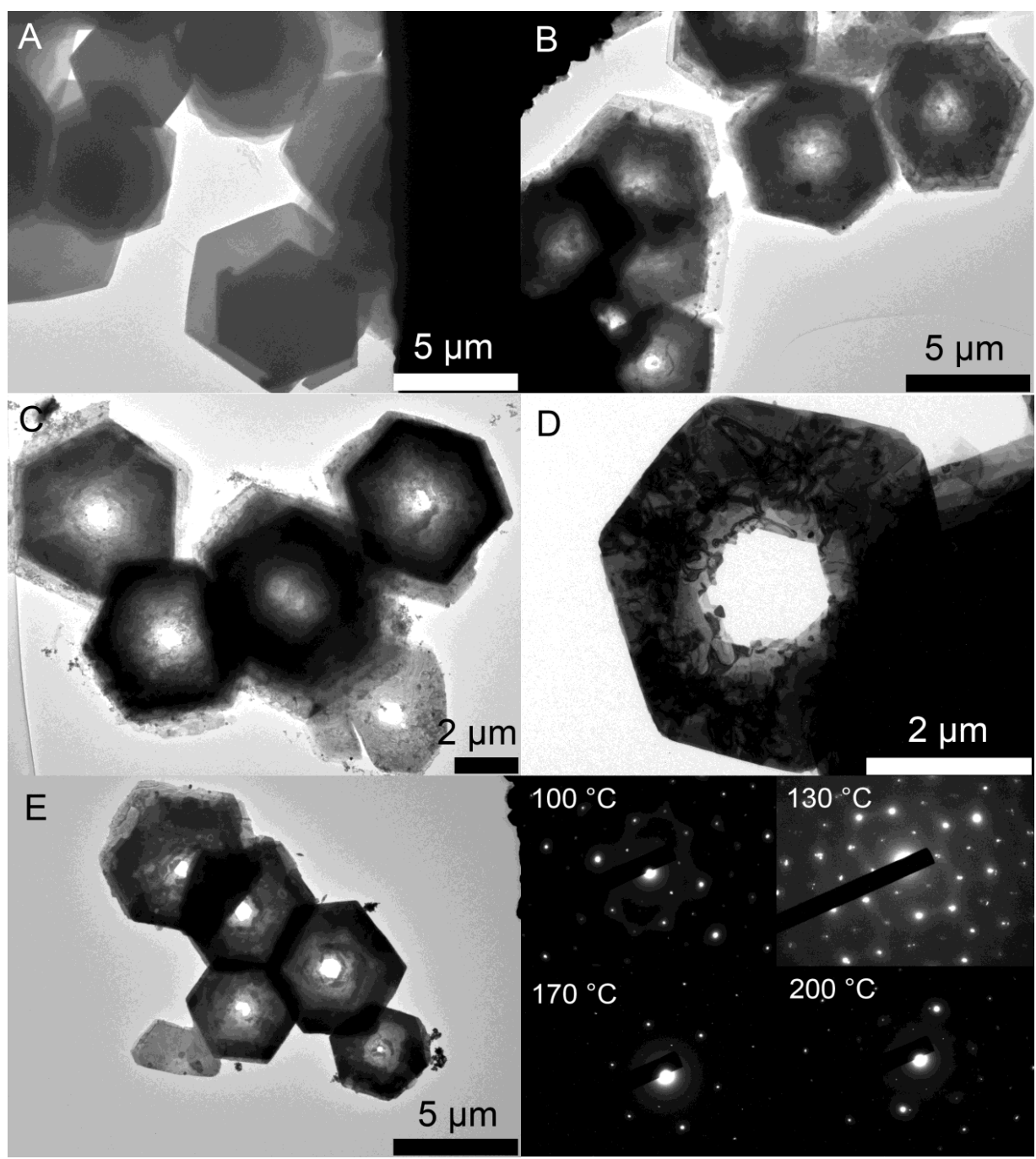

**Figure 1.** (A) TEM image of nanosheets prepared at 80 °C. (B) TEM image of rings prepared at 100 °C. (C), (D) and (E) rings synthesized at 130 °C, 170 °C and 200 °C, respectively. (F) SAEDs of all four samples.

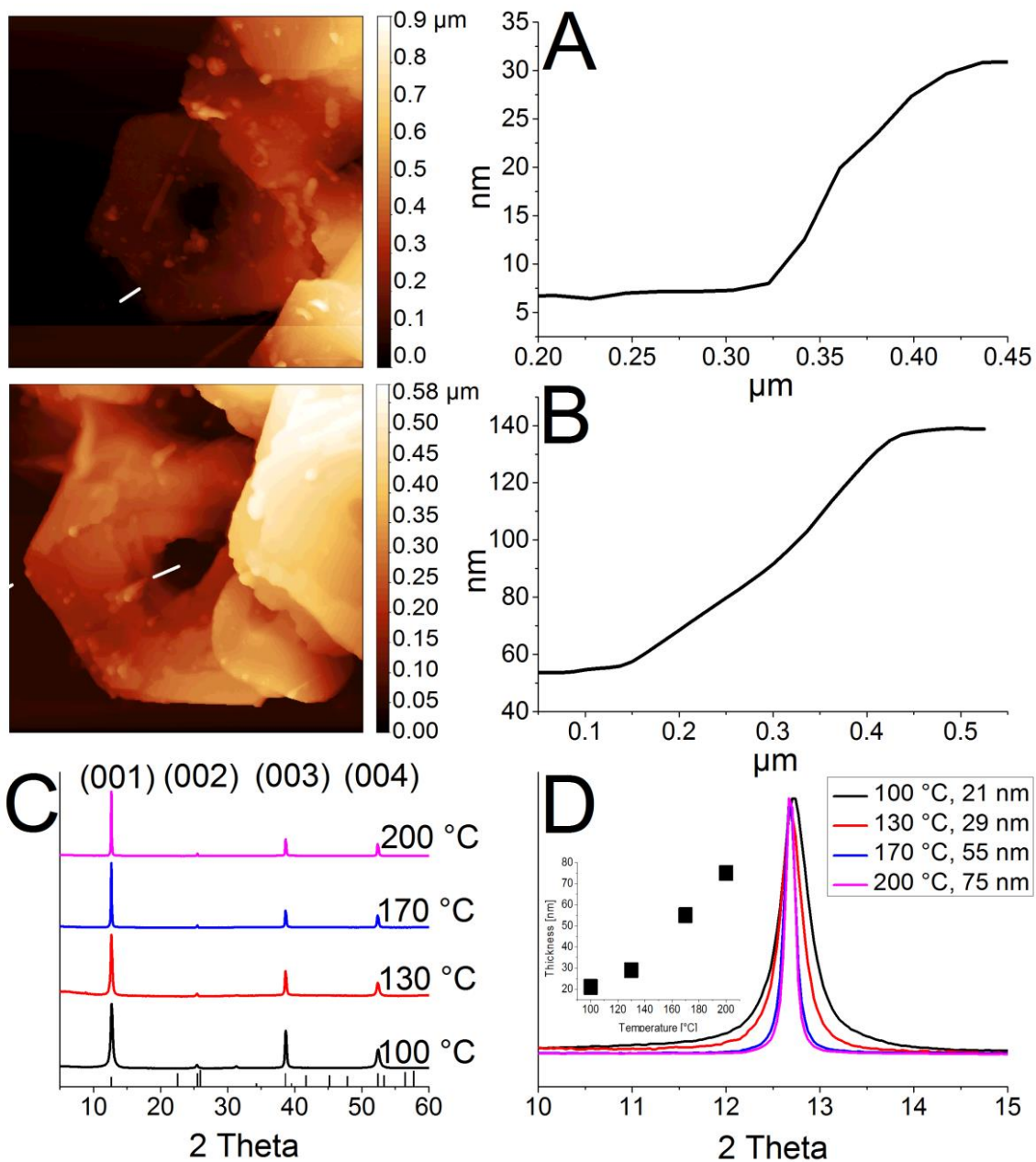

**Figure 2.** (A), (B) AFM images for rings synthesized at 100 °C and 200 °C. (C) Powder XRDs for samples prepared at different temperatures of 100 °C, 130 °C, 170 °C and 200 °C. (D) Change in thickness by varying the temperature.

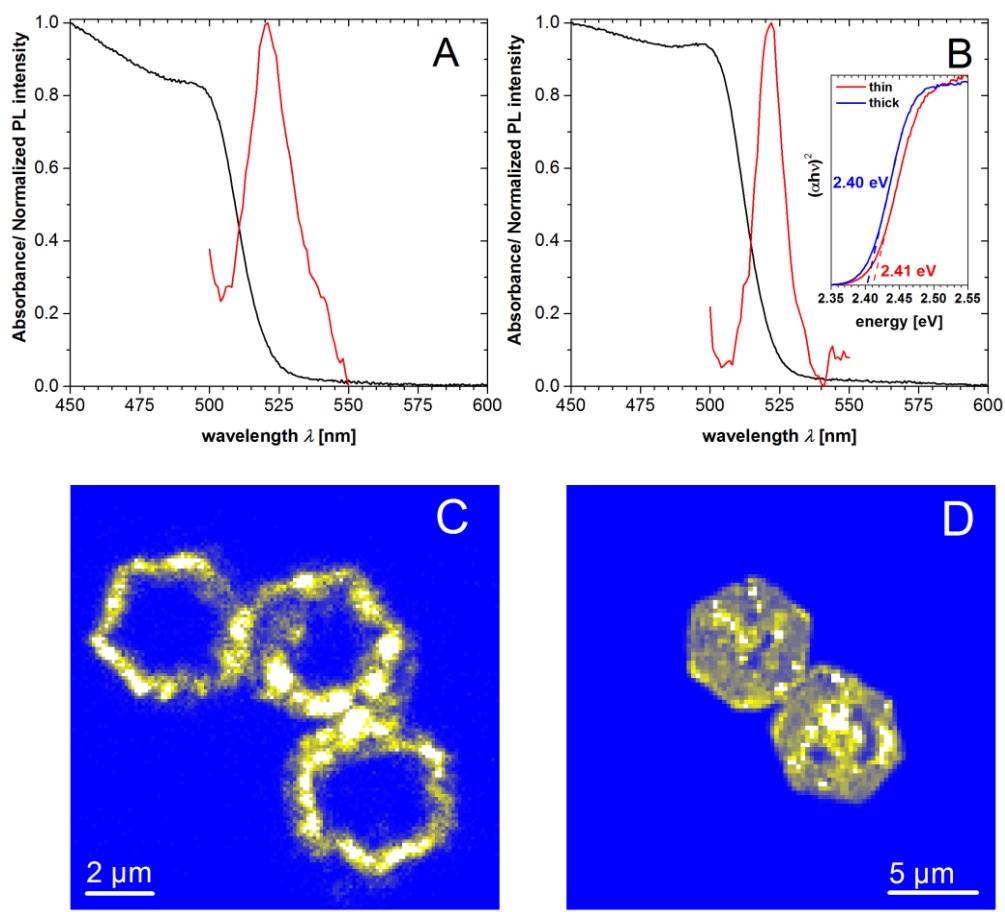

**Figure 3.** UV-Vis absorption (black) and emission (red) spectra of thin nanorings of 21 nm (A) and thicker nanorings of 82 nm (B) in toluene. Inset: Tauc plots for the absorption edge. (C), (D) PL mapping of individual rings by confocal excitation microscope. Presented rings belong to the samples with average thicknesses of 21 nm (C) and 82 nm (D) respectively. PL spectra of the individual structures can be found in Figure S3, E.

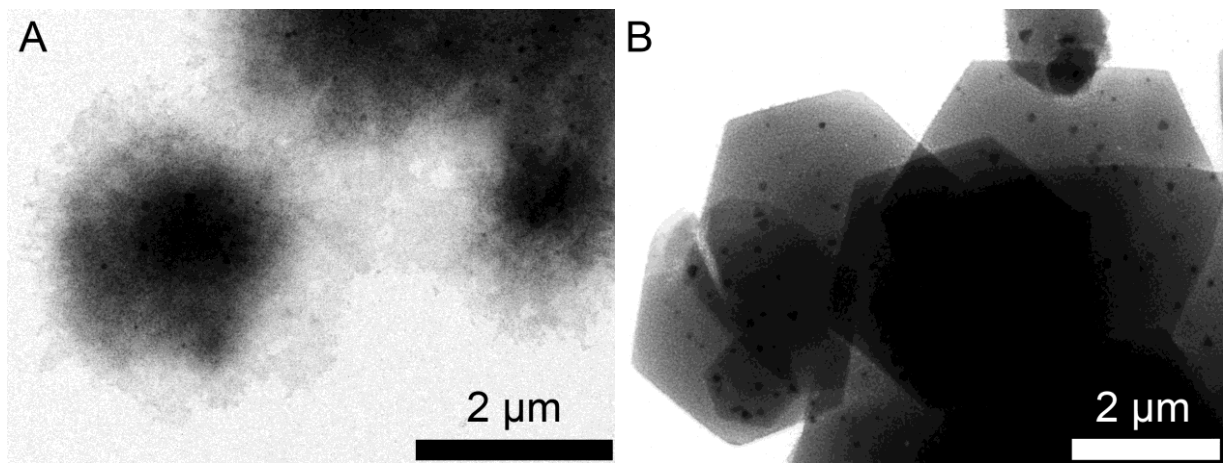

**Figure 4.** TEM images of the construction of a PbI$_2$ nanosheet in oleic acid. (A) Agglomeration of lead oleate and small PbI$_2$ particles. (B) Complete monocrystalline nanosheets at the end of the synthesis.

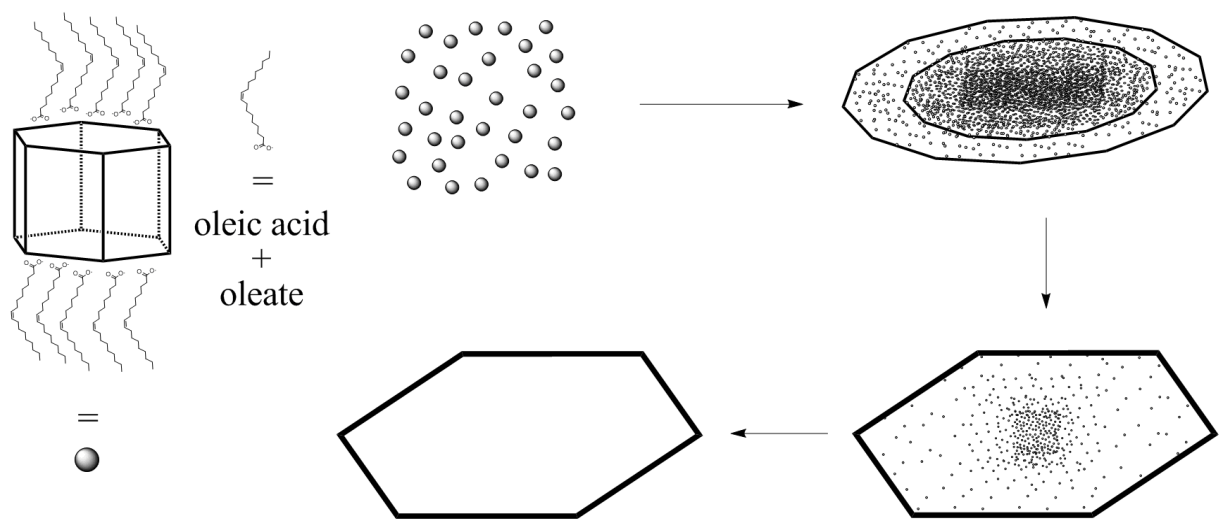

**Scheme 1.** Proposed mechanism for the formation of PbI$_2$ nanosheets following the preperation of PbI$_2$ particles which agglomerate to a sheet like shape. These particles crystallize with the time and form monocrystalline nanosheets.

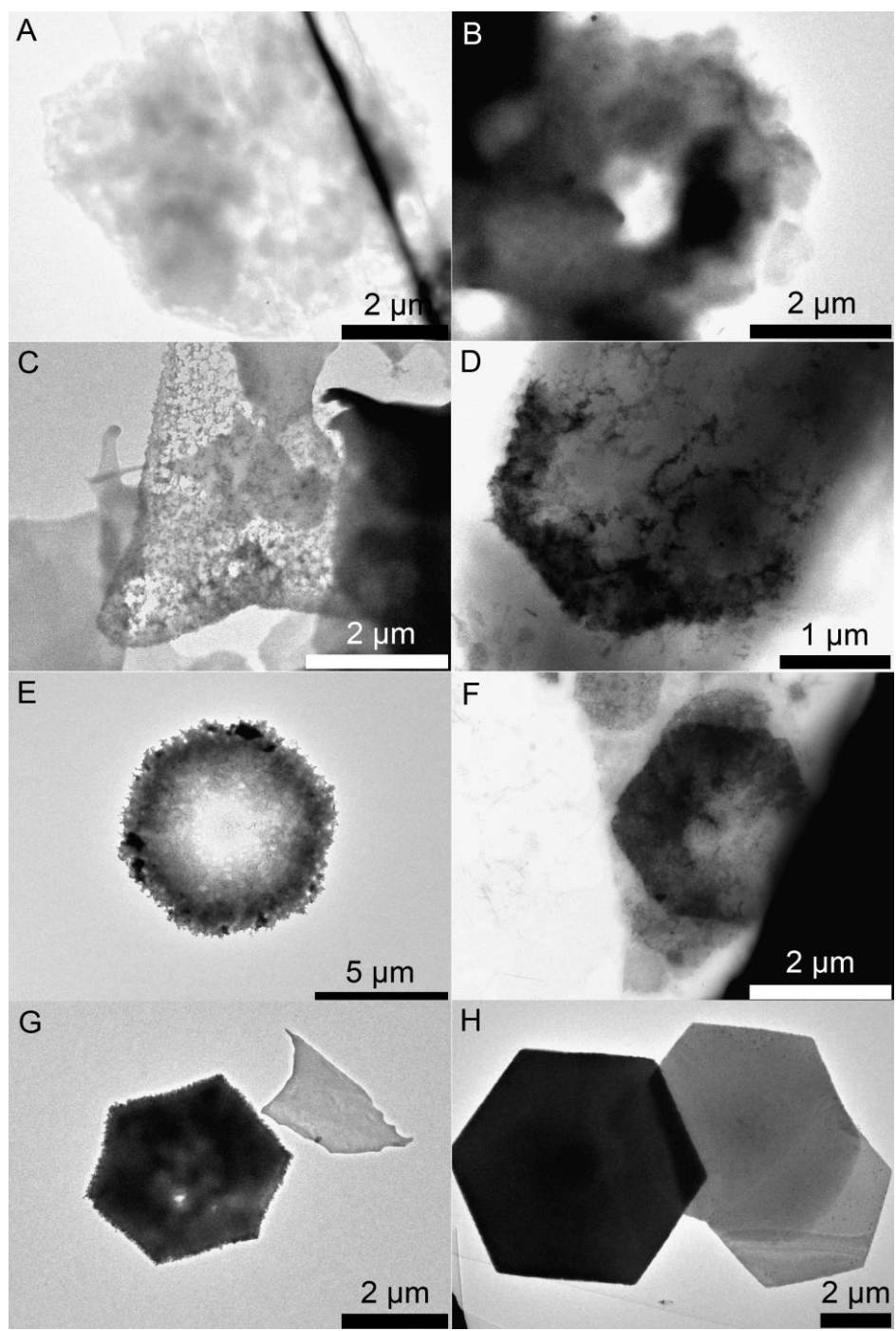

**Figure 5.** TEM images of the construction of a PbI$_2$ nanosheet in nonanoic acid. (A) Nonanoic acid and lead nonanoate forming a cloud. (B) Nonanoic acid and lead nonanoate forming a ring. (C), (D) Crystallization starts at the edges. (E), (F) Further growing of the shell. (G) A near complete sheet with a small hole in the middle. (H) Complete crystallization of the nanosheets.

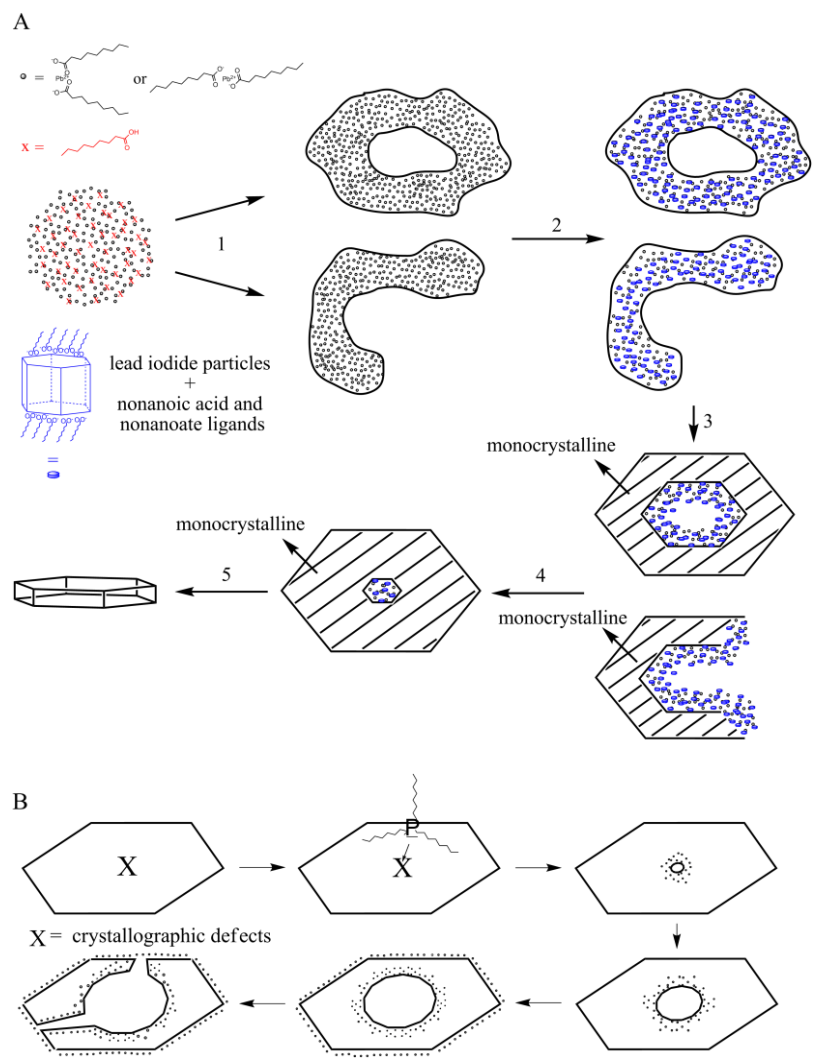

**Scheme 2.** Proposed mechanism for the formation of PbI$_2$ nanorings. (A) depicts the formation of PbI$_2$ nanosheets in nonanoic acid. At first lead nonanoate and nonanoic acid form a cloud. Step 1 depicts the separation of the lead oleate and the acid. To reduce the distance between the lead oleate molecules and therefore increase the reactivity nonanoic acid is pushed out of the cloud. Due to the large size of the cloud and the steric hindrance of the molecules the acid is pushed in two directions, outside and in the middle of the cloud. The result is the formation of ring or half ring clouds composed of lead nonanoate. In step 2 iodide ions react with lead to lead iodide particles. In step 3 lead iodide particles partly merge and form a monocrystalline ring while inside the lead nonanoate reacts further to lead iodide particles. At the end the hole is closed and single crystalline nanosheets of lead iodide are formed. (B) shows the etching of these sheets by TOP. At first only the middle is decomposed due to crystallographic defects. Later the sheet is etched everywhere. High amounts of TOP can destroy the sheets completely.

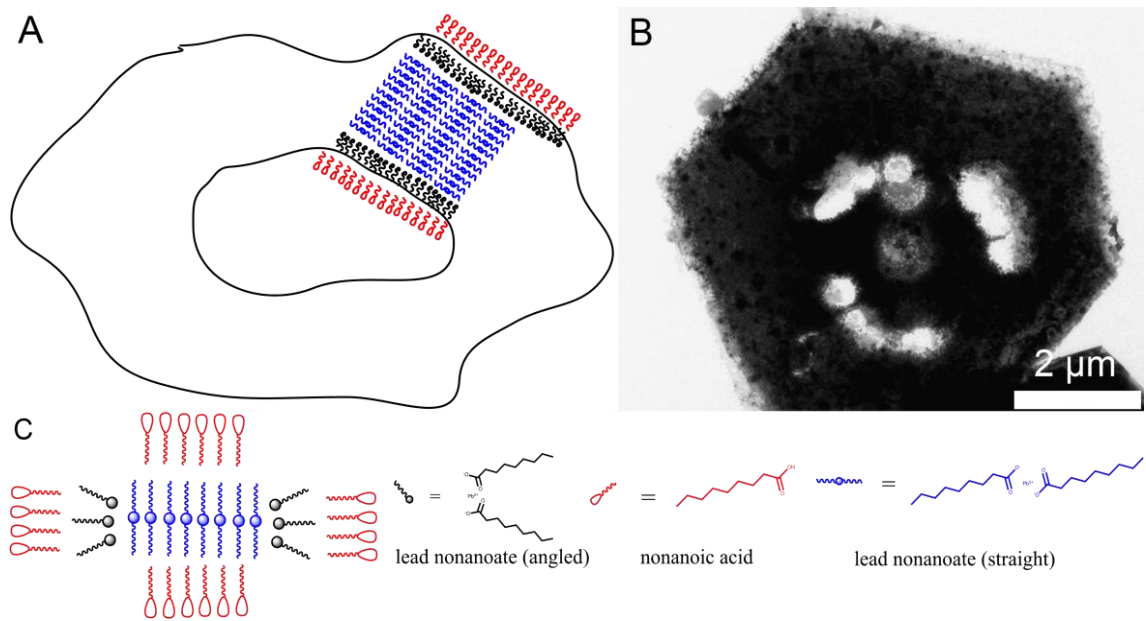

**Figure 6.** (A) depicts a schematic illustration of the proposed nature of the ring cloud. Inside the ring structure is composed of only lead nonanoate and outside only of nonanoic acid. (B) shows an etched nanosheet where the decomposition started not only in the center but at various other locations in the sheet. (C) Side view of the ring cloud together with the involved ligands in detail.

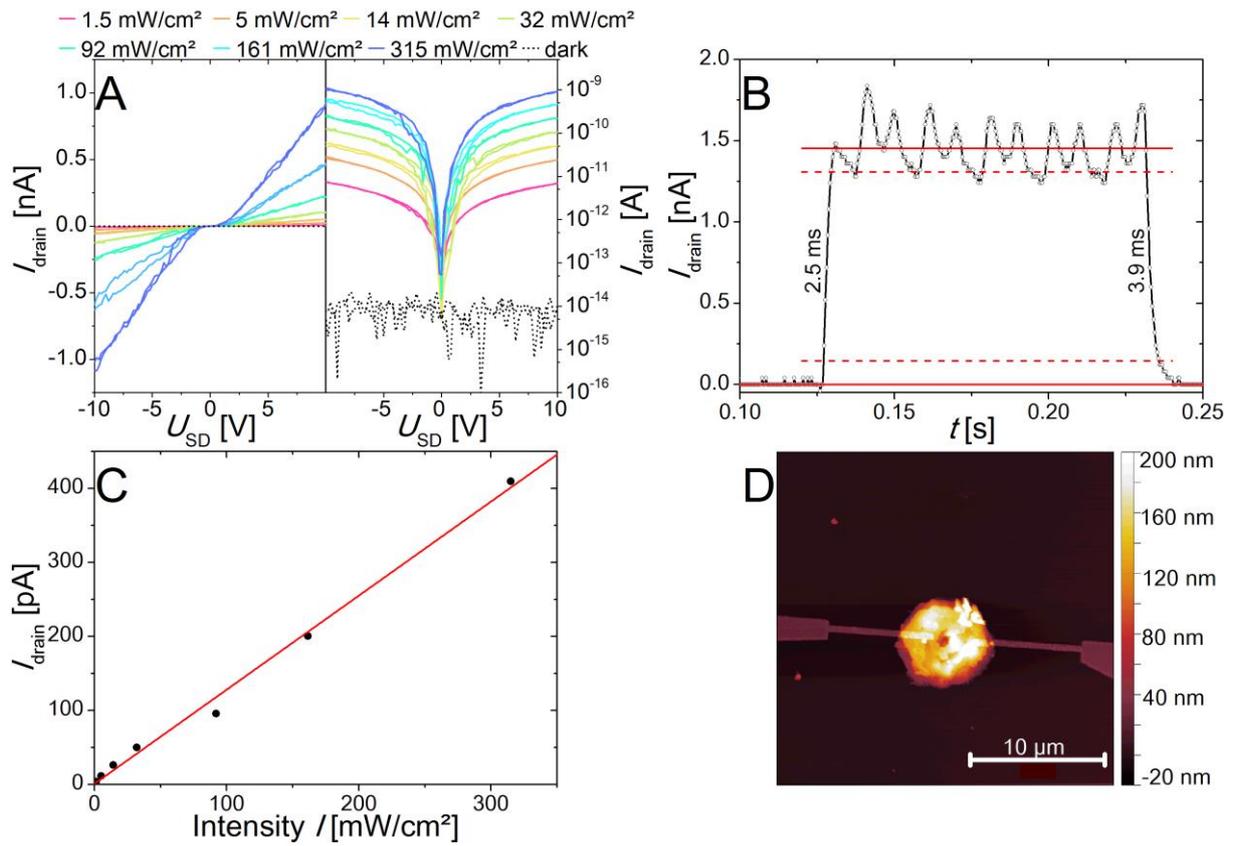

**Figure 7.** IV curves of a single PbI$_2$ nanoring device under illumination at different intensities at 405 nm in linear scale (left) and logarithmic scale (right) (A). Evaluation of rise and fall time of PbI$_2$ nanorings exposed by a pulsed 405 nm laser diode with 315 mW/cm² at a frequency of 50 Hz (B). The photocurrent at 5 V shows a linear dependency of the light power intensity used for excitation (C). AFM image of the characterized device (D).

**Table 1.** Figure of merits of the measured lead (II) iodide nanorings and other devices based on colloidal synthesis from the literature as well as graphene for comparison. The presented parameters were recorded for the devices shown in Figure 7 (82 nm thick nanorings) and Figure S10 (21 nm thick nanorings). Average numbers of devices prepared from nanosheets of the same thickness are given in brackets.

| | Response time rise/ fall [ms] | Gain | Responsivity [mA/W] | External quantum efficiency EQE [%] | Specific detectivity D* [Jones] | reference |
|---|---|---|---|---|---|---|
| Graphene | ps range | 1.25 | 1 | 6-16 | | 60 |
| $In_2S_3$ | 2000 / 100 | 155 | / | / | / | 57 |
| Cu doped $In_2S_3$ | 100 / 100 | 16 | / | / | / | 57 |
| $CsPbBr_3$ | 17.8 / 14.7 and 15.2 | $1·10^2<$ | / | / | / | 58 |
| GeS | 110 / 680 | / | $173·10^3$ | $5.32·10^4$ | $1.74·10^{13}$ | 59 |
| GeSe | 150 / 270 | / | $870·10^3$ | $2.67·10^5$ | $1.12·10^{13}$ | 59 |
| $PbI_2$ nanosheets | 20 | / | 1.3 | / | / | 61 |
| 82 nm thick $PbI_2$ nanorings | 2.5 / 3.9 (14.3/16.4) | $114·10^3$ ($35·10^3$) | 13.8 (8.4) | 4.2 (2.6) | $4.84·10^9$ ($2.70·10^9$) | This work |
| 21 nm thick $PbI_2$ nanorings | 12.0 / 31.1 (21.5/35.0) | $1.5·10^3$ ($2.2·10^3$) | 0.8 (0.9) | 0.25 (0.29) | $2.08·10^8$ ($2.52·10^8$) | This work |